\documentclass{iopjournal}

\usepackage{siunitx}
\usepackage{subfig}
\usepackage{amsmath}
\usepackage[switch]{lineno}
\usepackage{orcidlink}
\begin{document}

\articletype{Regular Article} %	 e.g. Paper, Letter, Topical Review...

\title{Record accumulation of antiprotons in a Penning-Malmberg Trap and their preparation for improved production of antihydrogen beams}

\author{
B.~Lee$^{1}$\orcidlink{0000-0002-7293-5142},
B.~Kim$^{2,*}$\orcidlink{0000-0003-0958-5503},
P.~Adrich$^{3}$\orcidlink{0000-0002-7019-5451},
I.~Belosevic$^{4}$\orcidlink{0000-0002-8001-8889},
M.~Chung$^{5}$\orcidlink{0000-0001-7014-4120},
P.~Comini$^{4}$\orcidlink{0000-0002-6373-4752},
P.~Crivelli$^{6}$\orcidlink{0000-0001-5430-9394},
P.~Debu$^{4}$\orcidlink{0000-0003-2988-5052},
S.~Geffroy$^{7}$\orcidlink{0009-0009-6940-8120},
P.~Guichard$^{8}$\orcidlink{0009-0003-8568-5327},
P.-A.~Hervieux$^{8}$\orcidlink{0000-0002-4965-9709},
L.~Hilico$^{9}$\orcidlink{0000-0002-8916-1294},
P.~Indelicato$^{9}$\orcidlink{0000-0003-4668-8958},
S.~Jonsell$^{10}$\orcidlink{0000-0003-4969-1714},
S.~Kim$^{1}$\orcidlink{0000-0002-0013-0775},
E.-S.~Kim$^{11}$\orcidlink{0000-0001-5603-6764},
N.~Kuroda$^{12}$\orcidlink{0000-0003-2727-790X},
L.~Liszkay$^{4}$\orcidlink{0000-0003-4371-4380},
D.~Lunney$^{7}$\orcidlink{0000-0002-3227-305X},
G.~Manfredi$^{8}$\orcidlink{0000-0002-5214-8707},
B.~Mansoulié$^{4}$\orcidlink{0000-0001-5945-5518},
M.~Matusiak$^{3}$\orcidlink{0000-0002-8239-6971},
V.~Nesvizhevsky$^{13}$\orcidlink{0000-0002-5364-0197},
F.~Nez$^{9}$\orcidlink{0000-0002-3478-7521},
K.~Park$^{1}$\orcidlink{0000-0002-6013-0259},
E.~Perez$^{14}$\orcidlink{0000-0002-7359-9689},
P.~Pérez$^{4}$\orcidlink{0000-0003-1407-1582},
C.~Regenfus$^{6}$\orcidlink{0000-0001-9656-3104},
C.~Roumegou$^{7}$\orcidlink{0009-0006-0343-9256},
J.-Y.~Roussé$^{4}$\orcidlink{0000-0000-0000-0000},
F.~Schmidt-Kaler$^{15}$\orcidlink{0000-0002-5697-2568},
K.~Szymczyk$^{3}$\orcidlink{0000-0001-9159-485X},
T.~A.~Tanaka$^{12, 17}$\orcidlink{0009-0003-8306-5546},
B.~Tuchming$^{4}$\orcidlink{0000-0002-1356-0723},
D.-P.~van~der~Werf$^{16}$\orcidlink{0000-0001-5436-5214},
D.~Won$^{1}$\orcidlink{0009-0005-5557-7709},
S.~Wronka$^{3}$\orcidlink{0000-0003-3277-138X},
P.~Yzombard$^{9}$\orcidlink{0000-0002-0864-181X}
}

\affil{$^{1}$Department of Physics and Astronomy, Seoul National University, Seoul, Korea}

\affil{$^{2}$Center for Underground Physics, Institute for Basic Science, Daejeon, Korea}

\affil{$^{3}$National Centre for Nuclear Research (NCBJ), ul. Andrzeja Soltana 7, 05-400 Otwock, Swierk, Poland}

\affil{$^{4}$IRFU, CEA, Université Paris-Saclay, F-91191 Gif-sur-Yvette, France}

\affil{$^{5}$The Pohang University of Science and Technology (POSTECH), Pohang, Republic of Korea}

\affil{$^{6}$Institute for Particle Physics and Astrophysics, ETH Zurich, 8093 Zurich, Switzerland}

\affil{$^{7}$Université Paris-Saclay, CNRS/IN2P3, IJCLab, Orsay, France}

\affil{$^{8}$Université de Strasbourg, CNRS, Institut de Physique et Chimie des Matériaux de Strasbourg, France}

\affil{$^{9}$Laboratoire Kastler Brossel, Sorbonne Université, CNRS, ENS-Université PSL, Collège de France, Paris, France}

\affil{$^{10}$Department of Physics, Stockholm University, Stockholm, Sweden}

\affil{$^{11}$Department of Accelerator Science, Korea University Sejong Campus, Sejong, Korea}

\affil{$^{12}$Institute of Physics, University of Tokyo, Tokyo, Japan}

\affil{$^{13}$Institut Max von Laue - Paul Langevin (ILL), Grenoble, France}

\affil{$^{14}$CERN, Geneva, Switzerland}

\affil{$^{15}$QUANTUM, Institut für Physik, Johannes Gutenberg Universität, Mainz, Germany}

\affil{$^{16}$Department of Physics, Swansea University, Swansea, United Kingdom}

\affil{$^{17}$Present address: National Metrology Institute of Japan (NMIJ), National Institute of Advanced Industrial Science and Technology (AIST), Tsukuba, Japan}

\affil{$^{*}$Author to whom any correspondence should be addressed.}

\email{name@institution.org}


% %%% Line Numbers for editors
% \linenumbers

\begin{abstract}
CERN's AD/ELENA ``antimatter factory'' - unique worldwide - serves several experiments, all of which use electromagnetic traps to accumulate antiprotons for fundamental science.
The GBAR experiment employs a charge-exchange reaction between an antiproton beam and a positronium cloud to produce antihydrogen for gravitational studies. 
% The GBAR experiment produces antihydrogen for gravitational studies by producing antihydrogen in flight through a charge-exchange reaction with a positronium cloud.
GBAR has also pioneered an electrostatic scheme using a pulsed drift tube to decelerate the 100 keV antiproton beam, rather than slowing the antiprotons in a foil, as is commonly done in other experiments. 
Following first results producing a 6 keV antihydrogen beam directly after the decelerator, a trap has now been installed to increase the production rate.  The emittance growth resulting from the deceleration is reduced in the trap by Coulomb interaction with a cold electron cloud.  The antiproton cloud is further compressed using rotating wall cooling and can be re-accelerated up to energies of 10 keV, including a time focus.  Here we describe the commissioning results, trapping 56(3)\% of the ELENA beam, delivering $6.4(0.4)~\times~10^{6}$ antiprotons per shot for improved production of antihydrogen, and a record accumulation of over $6.4(0.4)~\times~10^{7}$ antiprotons in under 35 minutes.
\end{abstract}

%%%%%%%%%%%%%%%%%%%%%%%%%%%%%%%%%%%%%%%%%%%%%%%%%%%%%%%%%%%%%
\section{Introduction}
\label{intro}
%%%%%%%%%%%%%%%%%%%%%%%%%%%%%%%%%%%%%%%%%%%%%%%%%%%%%%%%%%%%%
The GBAR experiment aims to measure the gravitational acceleration of antimatter by observing the free-fall of ultracold antihydrogen atoms in the terrestrial gravitational field~\cite{gbar_proposal,Perez2015}. Recent work proposes a method to achieve $10^{-6}$ precision in this measurement by employing a single-bounce quantum gravimeter~\cite{crepin19, guyomard25}.
The strategy for cooling antimatter to the 10~\si{\micro\kelvin} level is based on the production of positively charged antihydrogen ions ($\mathrm{\overline{H}}^+$), which can be confined in a Paul trap. By sympathetically cooling the $\mathrm{\overline{H}}^+$ ions with laser-cooled Be$^+$ ions, the production of ultracold antimatter would be feasible~\cite{gbar_origin, hilico2014}.
% Unlike neutral antihydrogen atoms, which are difficult to trap and cool, these positively charged ions can be confined in Paul traps. By sympathetically cooling the $\mathrm{\overline{H}}^+$ ions with laser-cooled ions in a Paul trap, the production of ultracold antimatter would be feasible~\cite{gbar_origin}.
Subsequently, an ultracold antihydrogen atom can be produced by photo-detachment of the excess positron.

The GBAR production scheme using charge exchange of antiprotons with a positronium (Ps) cloud is described in detail by Adrich et al. \cite{gbar_2022}.  
Antiprotons are provided by CERN's AD/ELENA facility \cite{AD_eff}, while the Ps cloud is created from positrons provided by a dedicated electron linac.  
The production of the antihydrogen ion requires a second charge-exchange reaction of the antihydrogen atom with the same Ps cloud.  

The production rate increases quadratically with the Ps density.  Since the initial production of antihydrogen using a flat Ps target \cite{gbar_2022}, we have installed a compact cavity.  While this increases the Ps density due to the smaller volume, it also places more stringent requirements on the emittance of the antiproton beam that must pass through it.  Therefore a means of reducing the antiproton-beam emittance is essential and best accomplished using a Penning-Malmberg trap, within which the antiprotons may be cooled by Coulomb interaction with cold electrons.

Achieving the high antiproton intensity is challenging, as the trapping efficiency in low-energy experiments using degrader foils has historically been constrained.
The collection of antiprotons into Penning-type traps was pioneered by
the PS196 collaboration at CERN's Low Energy Antiproton Ring (LEAR)~\cite{atrap0} and subsequently refined using the Antiproton Decelerator (AD) facility~\cite{atrap1}.
A fundamental improvement was realized with the commissioning of the ELENA project, which further cools and decelerates the AD beam to 100~\si{\kilo\electronvolt} with 80\% efficiency~\cite{elena}. Now, all experiments studying antimatter resort to Penning traps. 
The ASACUSA experiment replaced its former radiofrequency quadrupole decelerator (RFQD) setup~\cite{Kuroda2012} with a pulsed-drift-tube system and thin foil, demonstrating a trapping efficiency of $26(6)\%$ with $1.4(0.2) \times 10^{6}$ antiprotons per pulse~\cite{asacusa_2024} from ELENA.

The GBAR decelerator scheme consists of a drift-tube-electrode with fast high-voltage switching to reduce the 100~\si{\kilo\electronvolt} beam from ELENA to below 6~\si{\kilo\electronvolt}~\cite{husson21}, a configuration designed to achieve high transport efficiency. The antiprotons are subsequently captured and cooled within the trap via sympathetic cooling with pre-loaded electrons maintained at low temperature through synchrotron radiation. Following the cooling process, the trap facilitates the re-acceleration of the antiprotons up to 10~\si{\kilo\electronvolt} for further transport and subsequent antihydrogen production.

This paper describes the GBAR beam line optimized for the trapping and re-acceleration of antiprotons from ELENA and presents the first results of this high-efficiency scheme.

%%%%%%%%%%%%%%%%%%%%%%%%%%%%%%%%%%%%%%%%%%%%%%%%%%%%%%%%%%%%%
\section{Experimental setup}
\label{sec:1}
%%%%%%%%%%%%%%%%%%%%%%%%%%%%%%%%%%%%%%%%%%%%%%%%%%%%%%%%%%%%%
%%%%%%%%%%%%%%%%%%%%%%%%%%%%%%%%%%%%%%%%%%%%%%%%%%%%%%%%%%%%%
\subsection{Overview}

\begin{figure}
 \centering
        \includegraphics[width=\textwidth]{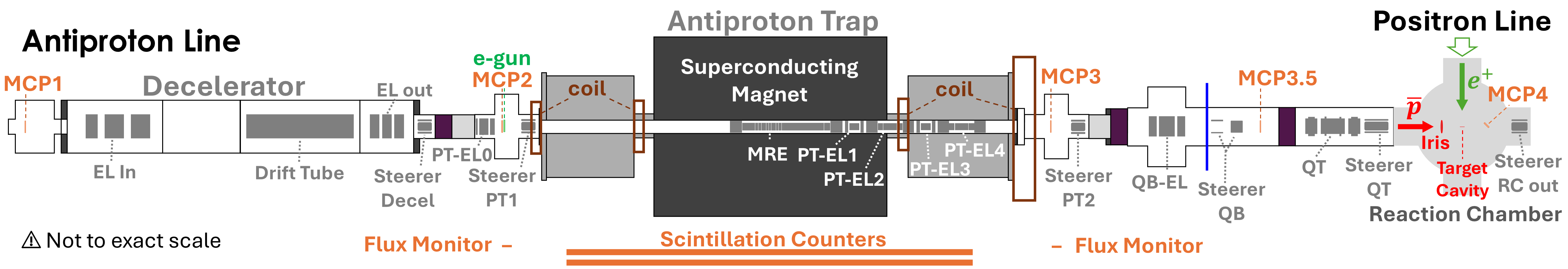}
 \caption{Schematic layout of the GBAR antiproton beamline.
The antiproton beam ($\mathrm{\overline{p}}$) from ELENA enters from the left, is slowed through the GBAR decelerator and transported to the antiproton trap ($\bar{p}$ trap, PT) for cooling, before extraction and transport to the reaction chamber. The assembly incorporates an electron gun (e-gun), magnetic guiding coils and various electrostatic beam optics, including Einzel lenses (EL), steerers, and a quadrupole triplet (QT).  Microchannel-plate (MCP) detectors, flux monitors, and a pair of scintillation counters are also installed for diagnostics.}
\label{fig:overview}
\end{figure}

The GBAR experiment consists of two orthogonal beam lines that intersect at the Ps target.  
In the positron beam line, 
a 9 MeV electron linear accelerator creates positrons with a tungsten target~\cite{positron2020}. These positrons are then cooled in a gas-filled, Surko-type trap before accumulation in a high-field Penning-Malmberg trap~\cite{positron2022}. The positron bunch is accelerated onto the porous silicon cavity, forming a dense positronium cloud through which passes the antiproton beam pulse. The neutral antihydrogen atoms that are produced travel straight through an electrostatic switchyard, 
designed to deflect any antihydrogen ions into another beam line while the unreacted antiprotons are deflected in the opposite direction.  

Fig. \ref{fig:overview} illustrates the antiproton beam line to the reaction chamber. Every 110~s, a pulsed antiproton beam with an energy of 100~keV is delivered from ELENA and decelerated to below 10~keV by the GBAR decelerator (described in Sec.~\ref{decel}).
The decelerated beam is then transported to the antiproton trap, where the particles are confined axially by an electric potential and radially by a uniform magnetic field (described in Sec.~\ref{trap}). 
After sufficient cooling, the antiproton cloud is transported to the reaction chamber electrostatically.
Stray magnetic fields of 100–300 $\mu$T measured along the beamline are compensated by steerers, while the reaction chamber is enclosed by iron plates to shield the interaction region from external magnetic fields.

The Ps target is a 20 mm long rectangular cavity with cross section of 1.5 mm $\times$ 2 mm, installed at the center of the reaction chamber concentric with the antiproton beam line. One side of the cavity features an entry window made of a 30 nm thick Si$_3$N$_4$ membrane to admit the incoming positron beam. Inside the cavity, a 1 $\mu$m thick mesoporous silica film is deposited on a conductive silicon substrate. Positrons strike this film to produce Ps which then escapes from the porous structure into the cavity volume.

The production of antihydrogen was previously demonstrated using a  target consisting of a simple plate that provided much lower Ps density, and a decelerated antiproton beam that was not trapped and cooled~\cite{gbar_2022}. 
When the ELENA beam is decelerated to energy $E_e$ from an initial energy $E_i$,  the transverse emittance increases by a factor of $\sqrt{E_i/E_e}$. This growth in emittance makes it challenging to focus the antiproton beam into the narrow target cavity. Therefore, it is essential to cool the beam within the antiproton trap after the deceleration process to achieve the beam quality required for efficient antihydrogen production.

%%%%%%%%%%%%%%%%%%%%%%%%%%%%%%%%%%%%%%%%%%%%%%%%%%%%%%%%%%%%%
\subsection{GBAR decelerator}
\label{decel}
The GBAR decelerator reduces the kinetic energy of the 100 keV antiproton beam from ELENA to below 10 keV for downstream trapping and reaction. This deceleration is based on the pulsed drift tube technique, where the electrode potential is switched to ground while the particles are inside the tube. Initial commissioning was performed in 2018 with a decelerated beam energy of 8 keV~\cite{husson21}. To ensure stable operation, the system was subsequently improved in 2023 with a redesigned electrode layout and a revised high-voltage switching circuit. These upgrades were essential to overcome corona and leakage currents in the negative high-voltage system and stringent ultra-high vacuum requirements~\cite{roumegou23}.

The GBAR decelerator consists of two Einzel lenses and a 450 mm long drift tube (Fig.~\ref{fig:overview}). The first Einzel lens is positioned to focus the beam as it enters the drift tube. This tube is typically held at -97 kV to reduce the beam energy to 3 keV, and is pulsed to ground using a fast high-voltage switch while the antiprotons are inside. A second Einzel lens and a four-plate steerer are located after the drift tube to correct the trajectory and ensure the decelerated beam is properly delivered into the antiproton trap. This pulsed drift tube design achieved nearly 100\% efficiency in decelerating antiprotons from ELENA and has been replicated by the PUMA experiment~\cite{fischer24}.

%%%%%%%%%%%%%%%%%%%%%%%%%%%%%%%%%%%%%%%%%%%%%%%%%%%%%%%%%%%%%
\subsection{Antiproton trap}
\label{trap}
\subsubsection{antiproton trap overview}

\begin{figure}
  \includegraphics[width=1\textwidth]{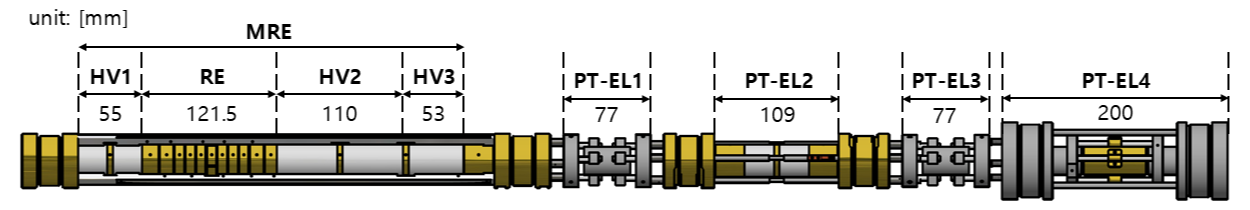}
\caption{
Detailed drawing of the electrode array within the antiproton trap, with dimensions %provided 
in mm. The Multi-Ring Electrode (MRE) assembly consists of a central RE section for the confinement of antiprotons and electrons, and high-voltage electrodes (HV1-3) dedicated to antiproton capture and acceleration. To ensure efficient beam transport, a series of four Einzel lenses (PT-EL1-4) are positioned downstream to focus and deliver the beam toward the reaction chamber.
}
\label{fig:trap}
\end{figure}

\begin{figure}
\centering
  \includegraphics[width=0.95\textwidth]{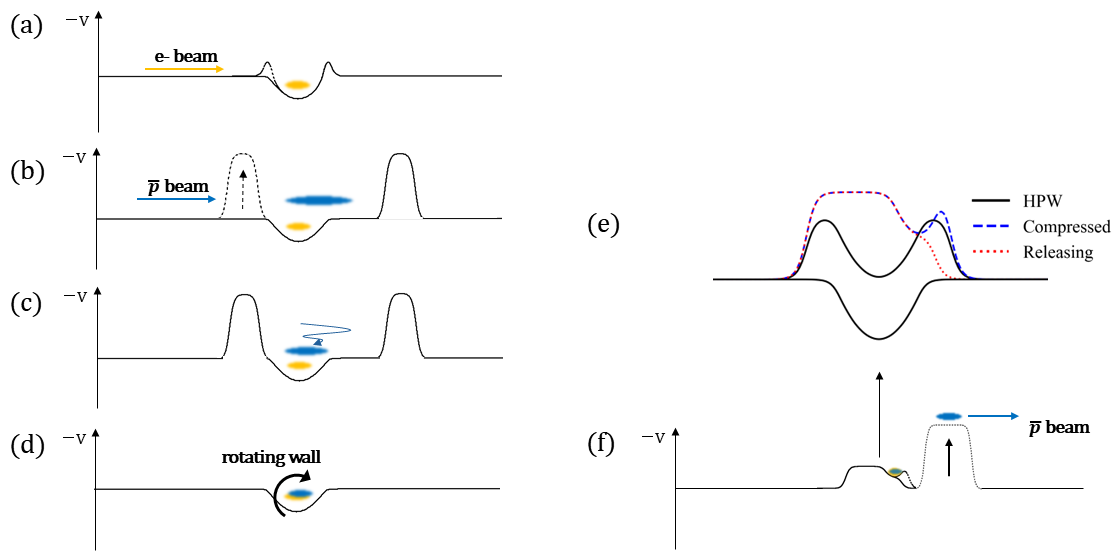}
\caption{Voltages along the trapping axis illustrating the operational sequence: (a) electron pre-loading, (b) antiproton capture, switching the HV1 electrode, (c) sympathetic cooling of antiprotons with cold electrons, (d) rotating wall compression, (e) extraction, and (f) acceleration to 1-10 keV. 
}
\label{fig:sequence}
\end{figure}

The antiproton trap is a Penning–Malmberg trap for the confinement of charged particles using electromagnetic static fields. It is designed to meet the requirements of the antiproton beam in the GBAR experiment, including a nested trap for electrons that cool the antiprotons. The overall operational schemes have been demonstrated through simulation studies~\cite{Yoo_2022}.
The antiproton trap consists of a cylindrical electrode array including Multi-Ring Electrodes (MRE) and a superconducting magnet.
The superconducting magnet can generate uniform magnetic fields up to 7 T, while the operational field is 5 T.
The MRE, as shown in Fig.~\ref{fig:trap}, consists of three High-Voltage electrodes (HV1, HV2, HV3) and an array of 11 Ring Electrodes (RE) between HV1 and HV2.
HV1 and HV3 are used to capture the antiproton beam by forming a high-voltage square potential well, and HV2 functions as a drift-tube accelerator for antiprotons.
RE form an electrostatic harmonic potential well (HPW) to confine and store cold antiprotons. The MRE is installed at the center of a cryo-cooled inner bore pipe vacuum chamber (IVC) inside the outer vacuum chamber (OVC) system with thermal shields, where a uniform and strong magnetic field is established by the superconducting magnet. A cryo-cooler system is mounted on the OVC chamber to cool the inner vacuum system, and a temperature of 15 K has been achieved around the MRE.

The voltages applied to the electrodes are controlled by a LabVIEW-based PXI system connected to amplifiers with a maximum output of $\pm$140 V and high-voltage power supplies with a maximum output of -20 kV. The central ring electrode is segmented into four pieces to apply a rotating wall signal.
The rotating wall technique~\cite{huang97, greaves00} is an established method for compressing non-neutral plasmas in a Penning trap by applying torque through rotating dipole electric fields.
Radial compression of antiproton clouds using this technique has been demonstrated through direct drive~\cite{Kuroda08}, sympathetic compression~\cite{Andresen08}, bounce-resonant transport~\cite{gutierrez15}, and multi-step high-density protocols~\cite{Aghion2018}.
For the last ring electrode, a fast switch is placed between the amplifier and the electrode to enable fast extraction of antiprotons.
The HV electrodes are connected to fast high-voltage switches. The switches for HV1 and HV3 are for antiproton capture, while the switch for HV2 is integrated into a double-gap buncher system~\cite{Yoo_2022}. This system consists of a high-voltage platform and a fast switch capable of generating a signal combining a time-varying potential and a fast high-voltage switching signal. It enables acceleration and bunch-length compression of the antiproton beam.
Downstream of the MRE, four Einzel lenses (PT-EL1-4) are installed to control beam divergence caused by the rapid magnetic field gradient.

\subsubsection{Antiproton trap operation}
The full sequence with operation functions of the antiproton trap is shown in Fig.~\ref{fig:sequence}.
First, a negative potential is applied to the RE to remove any residual ions from the previous sequence.
Then, the electron gun at the entrance of the antiproton trap is moved into the beam line and emits electrons into the trap with a current of 0.1-1 mA. 
The 20 eV electron beam is transferred adiabatically to the high magnetic field region using guiding coils.
A potential barrier of -140 V is formed by the last ring electrode, and a harmonic potential well is created before the potential barrier by the other ring electrodes. The continuous electron beam is reflected by the potential barrier, and some electrons lose their longitudinal energy through beam-beam interactions between incident and reflected beams, or through beam-plasma interactions with the confined electrons, eventually becoming trapped in the harmonic potential well~\cite{riken_etrap}. Using this scheme, approximately $3\times10^{8}$ electrons are accumulated within a few seconds.
After the electron accumulation, the electron gun is pulled out, and the PXI system waits for the ELENA ejection trigger signal. The confined electrons are cooled via cyclotron radiation in the high magnetic field region (Fig.~\ref{fig:sequence}a).
The antiproton beam is first decelerated to 3 keV by the drift tube decelerator and then delivered to the antiproton trap.
During injection, HV3 is held at –8 kV while HV1 is grounded, awaiting the arrival of the beam.
Once the entire bunch is contained between HV1 and HV3, the beam is captured by switching up the high voltage potential at HV1 (Fig.~\ref{fig:sequence}b).
The antiprotons are further decelerated and confined in the 140 V deep HPW through sympathetic cooling with the pre-loaded cold electron plasma for 5~s (Fig.~\ref{fig:sequence}c).
Then, the confined antiprotons are radially compressed by applying a rotating wall signal on the central azimuthally segmented ring electrode and maintained in this state for 1 min (Fig.~\ref{fig:sequence}d).
The harmonic potential well is then floated and axially compressed in 1 ms steps before the antiprotons are released by switching off the potential barrier established by the last ring electrode.
The extracted antiproton beam is transported to HV2, where it undergoes acceleration and bunching (Fig.~\ref{fig:sequence}e).
The HV2 functions as a drift tube accelerator by applying a fast high-voltage switching signal, accelerating the antiproton beam to 1-10 keV (Fig.~\ref{fig:sequence}f). Simultaneously, the double-gap buncher system applies a time-varying potential to reduce the bunch length of the beam.
The accelerated antiproton beam, with an energy of a few keV, experiences a strong magnetic field gradient of up to approximately 47~T/m as it exits the trap. A series of four Einzel lenses is used to recover the resulting divergence.

%%%%%%%%%%%%%%%%%%%%%%%%%%%%%%%%%%%%%%%%%%%%%%%%%%%%%%%%%%%%%
\subsection{Detector description}
\label{detector}

\begin{figure}
\centering
\subfloat[]{
  \includegraphics[width=0.55\textwidth]{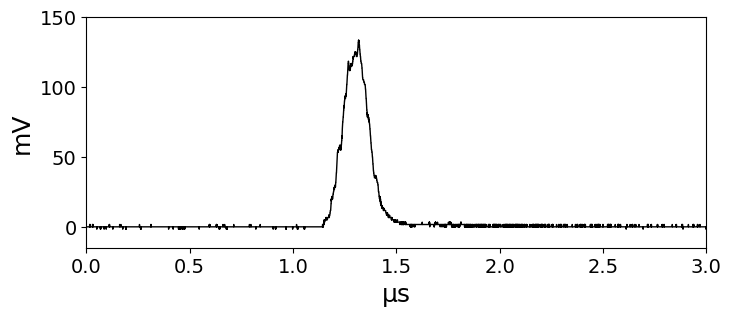}
}
\subfloat[]{
  \includegraphics[width=0.35\textwidth]{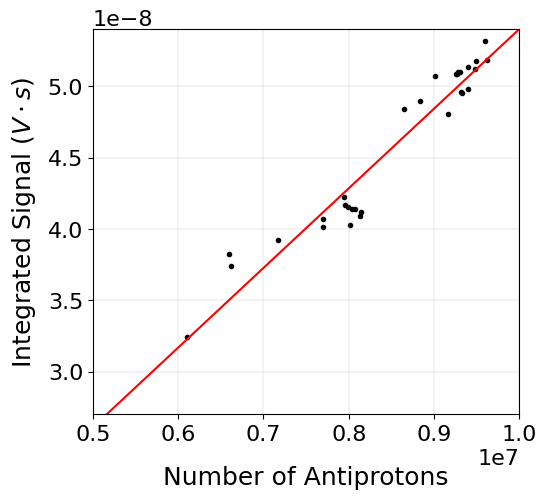}
}
\caption{
(a) Electric signal recorded by the flux monitor when a 100~keV antiproton beam is dumped onto MCP2, without using the GBAR decelerator. 
No upstream annihilation signal is observed prior to the beam arrival at MCP2.
(b) Correlation between the number of antiprotons and the integrated flux-monitor signal. The black dots represent the experimental data, and the red line indicates a linear fit. The data demonstrate a linear correlation with a deviation of approximately 3\%.
}
\label{fig:scintcali}
\end{figure}

Two types of detectors are used to monitor the status of the antiproton beam at several locations:  microchannel-plate (MCP) assemblies and plastic scintillators (Fig.~\ref{fig:overview}). Each MCP assembly consists of a microchannel plate, a phosphor screen, and a support structure. To capture the spatial distribution, CMOS cameras are positioned behind each assembly to record images of the phosphor light through a vacuum viewport. 
Simultaneously, the electrical signals from the MCPs are recorded by an oscilloscope to analyze the time distribution of the beam pulses.

While the MCP assemblies provide spatial and temporal information, plastic scintillation detectors are employed to monitor the beam intensity. These detectors are categorized into flux monitors and scintillation counters. Each flux monitor unit consists of a plastic scintillator (EJ-200, Eljen Technology) and a two-inch-diameter photomultiplier tube (PMT, H7195, Hamamatsu Photonics). The flux monitors employ scintillator plates with dimensions of $5 \times 7 \times 0.25$ cm$^3$ and are positioned at a distance of $2$~m from the beam line. They measure the antiproton flux by detecting secondary particles from the annihilation of the pulsed beam at MCP2 or MCP3, located at the entrance and exit of the antiproton trap, respectively. 

The scintillation counters comprise two 1.7 m long scintillator bars with area  10~cm $\times$ 5~cm, coupled to PMTs at both ends \cite{tof}. 
These counters are aligned parallel to the beam line at a  distance of $2.2$~m. While the flux monitors require dumping the antiproton beam onto the MCPs, the scintillation counters are designed to monitor the annihilation of the antiproton cloud during its confinement in the trap. A coincidence trigger condition is applied between the bars to selectively detect particles from the trap center.

The anode signals from both the flux monitors and scintillation counters are directed to a flash analog to digital converter (FADC) system integrated with a trigger and control board (TCB) for precise time synchronization and continuous monitoring, in parallel with an oscilloscope (Teledyne LeCroy Wavesurfer 3054z) for direct signal observation and analysis.

The flux monitors were calibrated relative to the antiproton pulse intensity measured by the ELENA nondestructive circular electrostatic pick-up system~\cite{pickup}.
This calibration assumes that the 100~keV antiproton pulses delivered by ELENA reach MCP2 without losses, as no upstream annihilation signal was observed prior to their arrival at MCP2 (Fig.~\ref{fig:scintcali}a). The resulting detector response demonstrated a linear correlation with the beam intensity, with a deviation of approximately $3\%$ (Fig.~\ref{fig:scintcali}b). The fitted intercept was comparable to the statistical deviation and thus consistent with zero.
By adjusting the high voltage of the PMTs, the integrated signal of the scintillator plates was matched to ensure a consistent response for the same antiproton pulse. After calibration, one monitor remained at MCP2, while the other was relocated to the MCP3 position, at the trap exit.

% In contrast, the scintillation counters underwent absolute calibration using cosmic muons because charged pions and muons exhibit similar energy deposition in the scintillator material.
% While twelve bars were employed for calibration, two were specifically selected as the detector, utilizing a six-bar coincidence trigger to select particles with small enough incident angles to ensure uniform energy deposition.

%%%%%%%%%%%%%%%%%%%%%%%%%%%%%%%%%%%%%%%%%%%%%%%%%%%%%%%%%%%%%
\section{Antiproton beam line commissioning results}
%%%%%%%%%%%%%%%%%%%%%%%%%%%%%%%%%%%%%%%%%%%%%%%%%%%%%%%%%%%%%
%%%%%%%%%%%%%%%%%%%%%%%%%%%%%%%%%%%%%%%%%%%%%%%%%%%%%%%%%%%%%
\subsection{Antiproton beam deceleration}

A bunched beam of $(1.0-1.2) \times 10^7$ antiprotons with an energy of 100 keV is delivered from ELENA every 110 seconds.
To minimize the bunch length, the bunch rotation method was applied at ELENA~\cite{elena}, reducing the bunch length ($\sigma$) to 40~ns. The beam emittance is 2 mm$\cdot$mrad~\cite{elena23}.

\begin{figure}[t]
\centering

\subfloat[]{
  \includegraphics[width=0.3\textwidth]{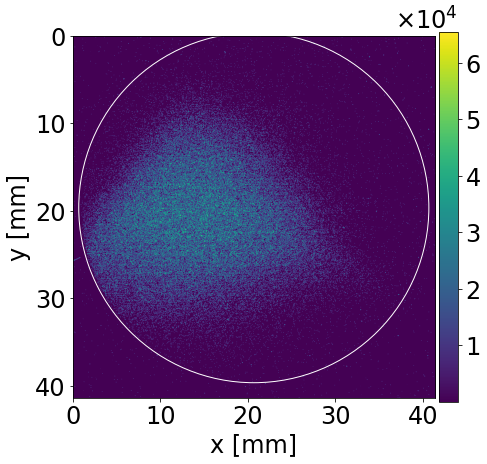}
}
\subfloat[]{
  \includegraphics[width=0.6\textwidth]{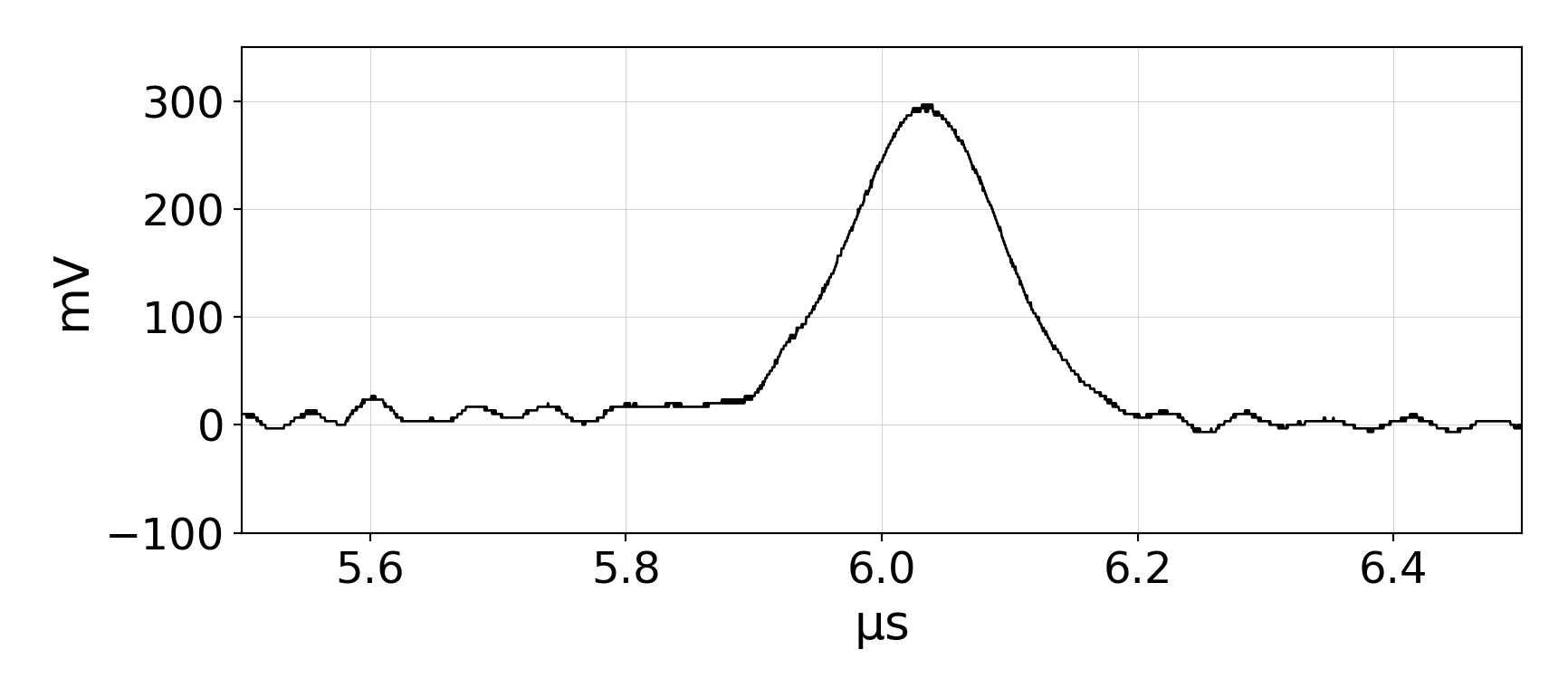}
}

\caption{
(a) Spatial profile and (b) electric signal for the 3~keV decelerated antiproton beam at MCP2. The circle in (a) indicates the active area of the MCP2. The bunch length in (b), defined as the full width at half maximum (FWHM) of the signal waveform, is 140~ns.
}
\label{fig:mcp2}
\end{figure}

The antiproton beam from ELENA is initially focused onto a drift tube using an Einzel lens (EL In in Fig.~\ref{fig:overview}) and then decelerated to below 10 keV.
After the deceleration, a second Einzel lens (EL out in Fig.~\ref{fig:overview}) focuses the beam toward the antiproton trap. The beam profile and the number of antiprotons are evaluated using MCP2. The corresponding spatial profile and electric signal for the antiproton beam decelerated to 3~keV are shown in Fig.~\ref{fig:mcp2}. The beam size, defined as the Gaussian fit standard deviation, is measured to be $\sigma_x$~=~5~mm, $\sigma_y$=~8~mm and the bunch length (FWHM) is 140~ns, and the number of decelerated antiprotons, as measured using the flux monitor, indicates a deceleration efficiency close to 100\%.
The 3 keV antiproton beam, routinely used for trapping, is expected to have an emittance $\sqrt{E_i/E_e}=\sqrt{100/3}\approx6$ times larger than that of the original ELENA beam due to the deceleration.

%%%%%%%%%%%%%%%%%%%%%%%%%%%%%%%%%%%%%%%%%%%%%%%%%%%%%%%%%%%%%
\subsection{Antiproton trapping and cooling}

The 3 keV antiproton beam is delivered to the antiproton trap with a slight reduction in transmission due to imperfect alignment  and the beam divergence.
In the ``pass-through'' mode, where the antiproton beam is transferred through the antiproton trap and dumped onto MCP3, 68(8)\% of the ELENA beam intensity, corresponding to $(7-8)\times 10^6 \bar{p}$ is detected at the exit of the trap. The antiproton beam intensity is estimated using the flux monitor described in section \ref{detector}.

The number of captured antiprotons is also measured using the flux monitor by releasing antiprotons after a short storage period of 10 $\mu$s and dumping them onto MCP3.
It is maximized by optimizing the switching time of HV1, adjusted in steps of 100 ns. 
The number of captured antiprotons is measured to be 81(10)\% of the number of particles in ``pass-through'' mode.
The capture efficiency is limited by the bunch length, the energy distribution and the angle of the injected antiproton beam.

\begin{figure}
\centering
\includegraphics[width=0.5\textwidth]{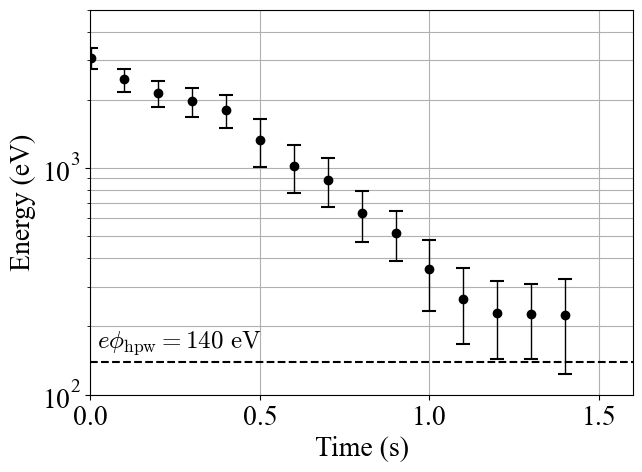}
\caption{The measured energy of the antiprotons
%at various cooling times. 
as a function of cooling time.
A horizontal dashed line indicates the energy threshold of 140 eV, below which the antiprotons 
%fall into and become 
are confined in the harmonic potential well.}
\label{fig:cooling_time}
\end{figure}

The captured antiprotons are slowed through interactions with the pre-loaded cold electron cloud. Once their energy drops below 140 eV, the antiprotons fall into the harmonic potential well and are confined with the electron plasma, where they are further cooled sympathetically.
The energy of the antiprotons before the confinement was measured by opening the potential barrier at HV3 and measuring their arrival time at MCP3.
As shown in Fig.~\ref{fig:cooling_time}, the antiproton energy decreases with time as they are sympathetically cooled by the preloaded electrons. However, since antiprotons confined in the harmonic potential well cannot escape the trap by opening the HV3 barrier, only particles with energies above 140 eV can escape 
%the potential well 
and be detected. After 1.5 s, no particles are detected when the potential barrier is opened, indicating that the entire antiproton population has become confined within the harmonic potential well.

%%%%%%%%%%%%%%%%%%%%%%%%%%%%%%%%%%%%%%%%%%%%%%%%%%%%%%%%%%%%%
\subsection{ Radial Compression of an Antiproton cloud }

\begin{figure}
\centering
\begin{minipage}[c]{0.48\textwidth}
\subfloat[]{
    \centering
    \includegraphics[width=\textwidth]{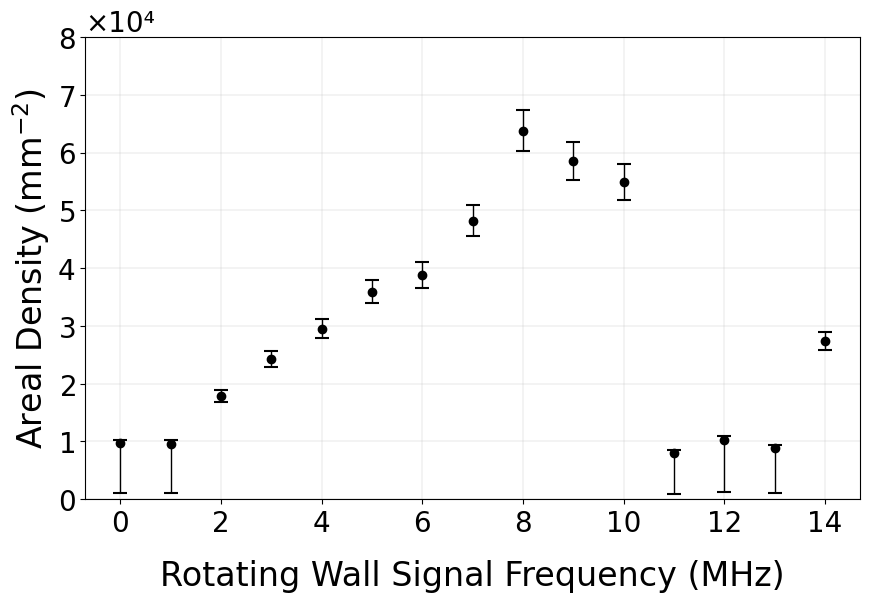}
    \label{fig:freq_density}
}
\end{minipage}
\hfill
\begin{minipage}[c]{0.48\textwidth}
    \centering
    \subfloat[OFF]{
        \includegraphics[width=0.48\textwidth]{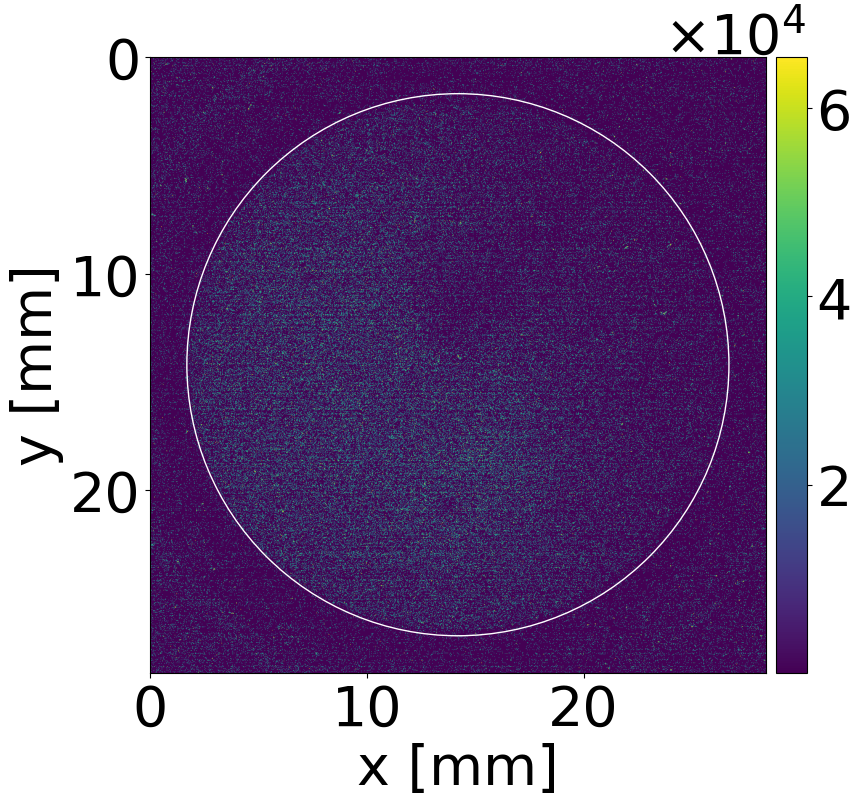}
        \label{fig:off}
    }
    \subfloat[4 MHz]{
        \includegraphics[width=0.48\textwidth]{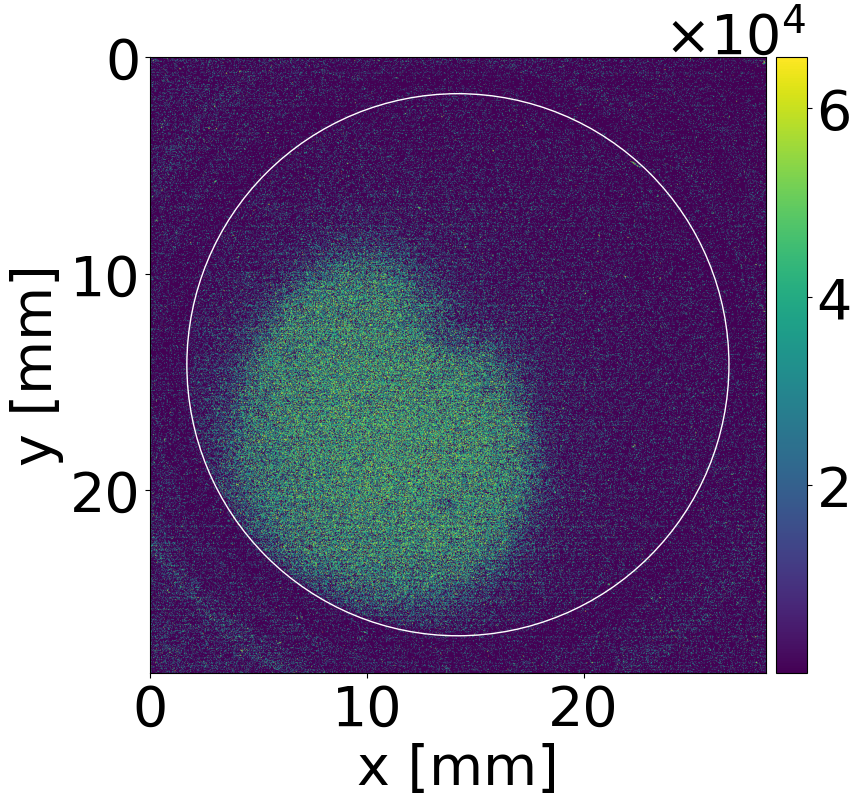}
        \label{fig:4mhz}
    }

    \subfloat[8 MHz]{
        \includegraphics[width=0.48\textwidth]{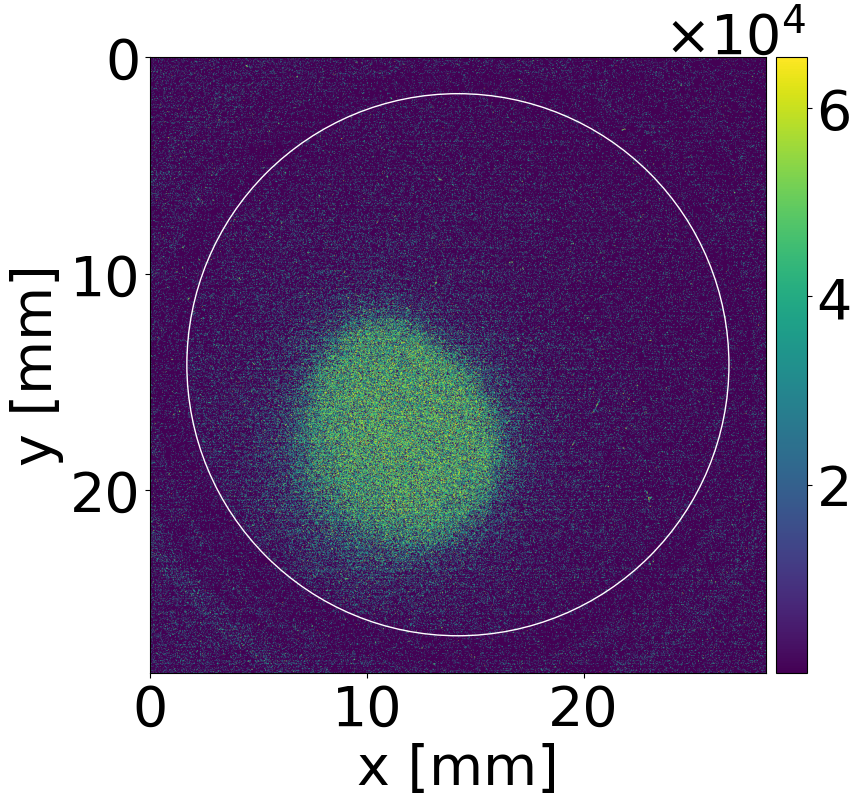}
        \label{fig:8mhz}
    }
    \subfloat[12 MHz]{
        \includegraphics[width=0.48\textwidth]{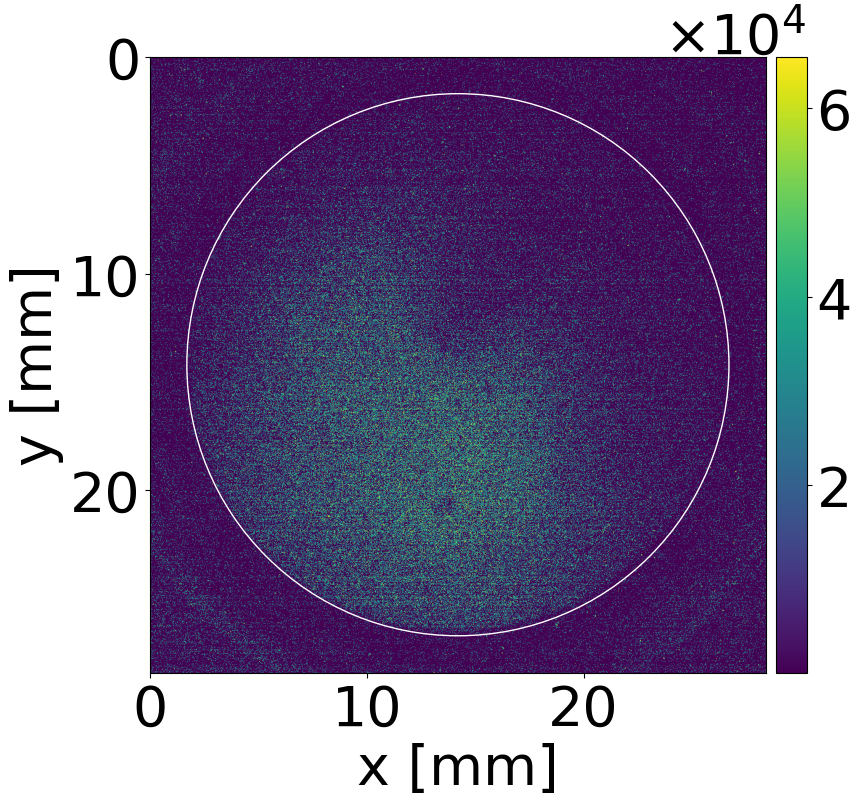}
        \label{fig:12mhz}
    }
\end{minipage}
\caption{Areal density (a) and spatial profiles (b--e) of the extracted antiproton beam as a function of rotating wall (RW) frequency ($\omega/2\pi$) in the range 1--14~MHz, where the amplitude is fixed at 10~V. The beam profiles at MCP3 are shown for (b) before compression and after RW compression at (c)~4~MHz, (d)~8~MHz, and (e)~12~MHz. The white circle indicates the active area of the MCP. The signal attenuation, appearing as a diminished intensity region slightly above the center of the image, is attributed to localized damage to the MCP.}
\label{fig:pbar_rw_scan}       % Give a unique label
\end{figure}

The segmented electrodes of the MRE, located at the center of the ring electrode assembly, are used to generate the rotating electric fields.
RF voltages with amplitudes of 1-10~V and frequencies up to 20~MHz can be applied in this configuration. Under these conditions, the electron plasma is expected in the ``strong-drive regime'' of rotating wall compression~\cite{fajans20}, where the plasma density increases linearly with RF frequency, provided that the plasma rotation matches the drive frequency. The antiproton cloud is then sympathetically compressed with the electron plasma. At higher frequencies, however, the plasma may fail to follow the drive, causing the density to deviate from this linear relationship.

The compressed antiprotons are extracted and delivered to MCP3, where the beam profile is recorded to evaluate the areal density of the extracted antiproton beam. The extraction process is described in the following section. The number of antiprotons reaching MCP3 is measured using a flux monitor. To estimate the beam area, an elliptical fitting algorithm is applied to the recorded image. The axis lengths are defined as the Full-Width at Half-Maximum (FWHM). The areal density is then calculated as the ratio of the number of antiprotons to the fitted beam area. The uncertainty in the areal density is determined by propagating the uncertainties in the number of antiprotons and the fitting error.
The optimal compression is studied by scanning the areal density while varying the rotating wall frequency ($\omega/2\pi$) in the range 1--14~MHz
% The optimal compression is studied by scanning the areal density with rotating wall frequency change in the range of $\omega/2\pi = $1-14~MHz
as shown in Fig.~\ref{fig:pbar_rw_scan}.
The areal density of the extracted antiproton beam reaches a maximum value at 8~MHz and drops at frequencies above 10~MHz. For drive frequencies above this value, sympathetic compression was observed to be ineffective. This behaviour is attributed to the reduced efficiency of the RW-driven compression of the electron plasma at higher frequencies. When the beam profile exceeds the effective area of MCP3, the fitting error in the beam size estimation becomes significant.

Rotating wall compression also helps to suppress plasma expansion, thereby increasing its lifetime~\cite{huang97}. Based on measurements of the remaining antiprotons, the lifetime of the antiproton plasma is estimated to exceed 10,000 seconds.

%%%%%%%%%%%%%%%%%%%%%%%%%%%%%%%%%%%%%%%%%%%%%%%%%%%%%%%%%%%%%
\subsection{Antiproton extraction, acceleration and bunching}

\begin{figure}[b]
\centering
\subfloat[]{
  \includegraphics[width=0.48\textwidth]{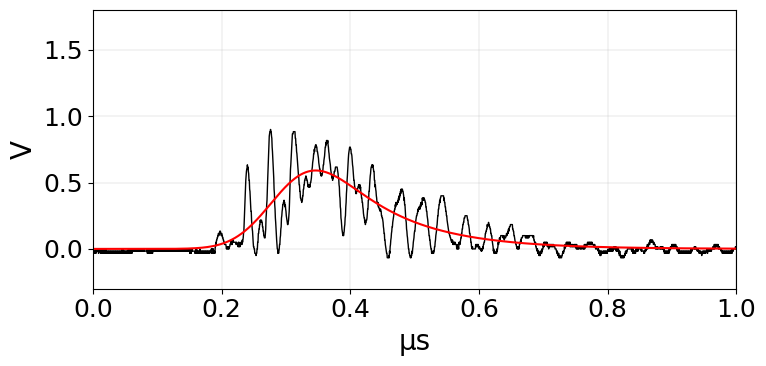}
}\hfill
\subfloat[]{
  \includegraphics[width=0.48\textwidth]{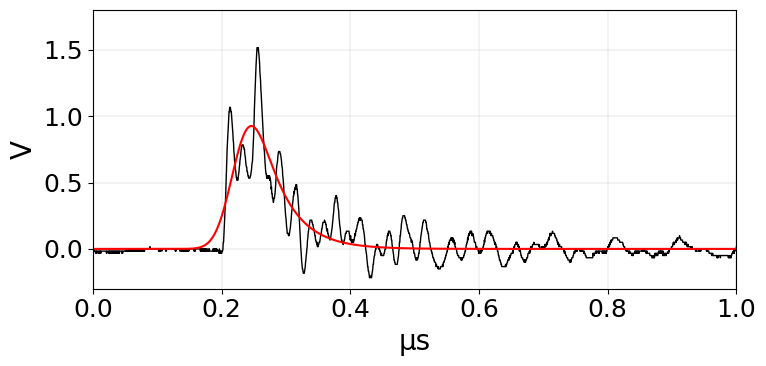}
}
\caption{
Extracted antiproton beam signals measured at MCP3, overlaid with an exponentially modified Gaussian (EMG) fit (red), for (a) Without a bunching signal and (b) with the double-gap buncher system active. The bunch length, defined as the full width at half maximum (FWHM) of the EMG fit, is reduced from 190~ns to 80~ns.}
\label{fig:pbar_buncher}       % Give a unique label
\end{figure}

The extraction of the antiprotons is performed by adjusting the potentials applied to the RE using the sequencing system. Firstly, the entire harmonic potential well is floated and axially compressed, with the bottom of the well set to $-70$~V.
The mixed plasma of electrons and antiprotons is then released using a fast switch connected to the last electrode in RE. The electrons quickly escape, while the antiprotons relatively move slowly toward the HV2, which is located just beyond the last ring electrode.
Once the antiprotons are fully contained within HV2, the potential in HV2 rapidly ramps up to the high-voltage potential and the antiprotons are accelerated by the pulsed drift tube technique.
The mean energy of the accelerated antiproton beam is estimated by measuring its time-of-flight to MCP3. The measured energy agrees with the applied voltage, with a deviation of less than 1\%.
Simultaneously, a double-gap buncher~\cite{kim2007, kim2012} potential is applied to compress the bunch length of the accelerated antiproton beam. Simulation studies have shown that the sawtooth-shaped bunching potential is expected to significantly reduce the bunch length~\cite{Yoo_2022}.
By applying this bunching signal, the bunch length (FWHM) of the extracted antiproton beam is reduced from 190~ns to 80~ns, as shown in Fig.~\ref{fig:pbar_buncher}. To determine the bunch length, the temporal profile of the beam pulse was fitted with an exponentially modified Gaussian (EMG) function~\cite{grushka1972}.
% \begin{center}
% $f(t; \mu, \sigma, \tau) = \frac{A}{2\tau} \exp \left[ \frac{1}{2} \left( \frac{\sigma}{\tau} \right)^2 - \frac{t - \mu}{\tau} \right] \text{erfc} \left( \frac{\mu - t}{\sqrt{2}\sigma} + \frac{\sigma}{\sqrt{2}\tau} \right)$
% \end{center}
% where $\mu$ and $\sigma$ represent the mean and standard deviation of the Gaussian component, respectively, and $\tau$ denotes the decay constant of the exponential tail.
The FWHM was subsequently extracted from the fitted EMG profile.

The extracted antiproton beam, typically with an energy of 6~keV, is focused by a series of four Einzel lenses positioned downstream of the MRE. These lenses prevent beam divergence caused by the magnetic field gradient at the exit of the trap, and focus the beam onto MCP3. The beam parameters are evaluated using MCP3 and the flux monitor near MCP3.
The number of antiprotons was measured to be $6.4(0.4) \times 10^6 $ on average, corresponding to $56(3)\%$ of the ELENA beam intensity.

%%%%%%%%%%%%%%%%%%%%%%%%%%%%%%%%%%%%%%%%%%%%%%%%%%%%%%%%%%%%%
\subsection{ Antiproton Accumulation }

\begin{figure}
\centering
  \includegraphics[width=0.48\textwidth]{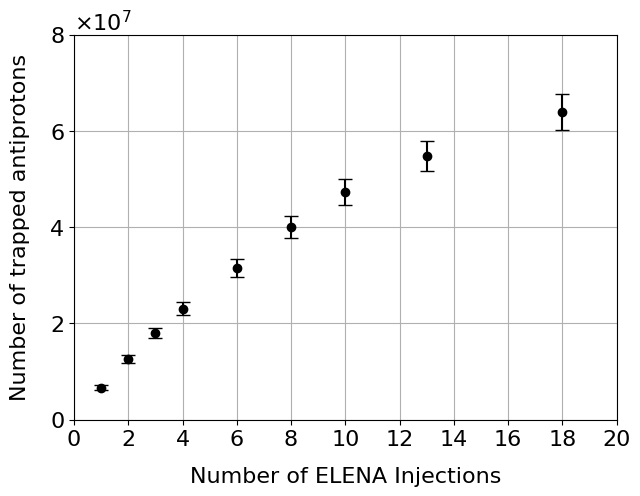}

\caption{
Accumulation of antiprotons in the antiproton trap as a function of the number of ELENA injections. The total number of confined antiprotons increases with subsequent stacking cycles, reaching a population of $6.4(0.4) \times 10^7$ after 18 injections. The number of antiprotons is measured by extracting and dumping the particles at MCP3 using the flux monitor.
} 
\label{fig:accumulation}
\end{figure}

In addition to the routine extraction of antiprotons after each 110~s ELENA cycle, the system can be operated in an accumulation mode where particles are retained to accumulate particles from subsequent ELENA injections. In this configuration, the antiproton trap functions as a reservoir, providing significant operational flexibility for the entire experimental sequence. The stacking procedure is carried out as follows: after the cooling and compression of the antiprotons, the cold antiprotons and electrons are maintained within the HPW. The RW compression is turned off before the subsequent injection to allow for expansion of the trapped particles, ensuring a sufficient spatial overlap with the incoming beam. High voltage is then applied to the HV3 electrode to form a potential wall to await the next ELENA injection. By repeating the capture, cooling, and compression cycles, a progressively larger population of antiprotons is confined within the HPW.

During these operations, approximately $6.7\times10^6$ antiprotons were trapped from the first ELENA injection, reaching a total of $6.4(0.4)\times10^7$ antiprotons after stacking 18 injections, as shown in Fig.~\ref{fig:accumulation}. This result marks a new record that significantly exceeds the $1\times10^7$ antiprotons previously reported by the ASACUSA experiment \cite{Kuroda2012}, while also demonstrating the highest accumulation rate to date by reaching the previous record's level in only two ELENA injections.

%%%%%%%%%%%%%%%%%%%%%%%%%%%%%%%%%%%%%%%%%%%%%%%%%%%%%%%%%%%%%
\section{Comparison to simulation} 
%%%%%%%%%%%%%%%%%%%%%%%%%%%%%%%%%%%%%%%%%%%%%%%%%%%%%%%%%%%%%

\begin{figure}[t]
\centering
\subfloat[\centering simulated beam profile at MCP3]{%
  \includegraphics[width=0.28\textwidth]{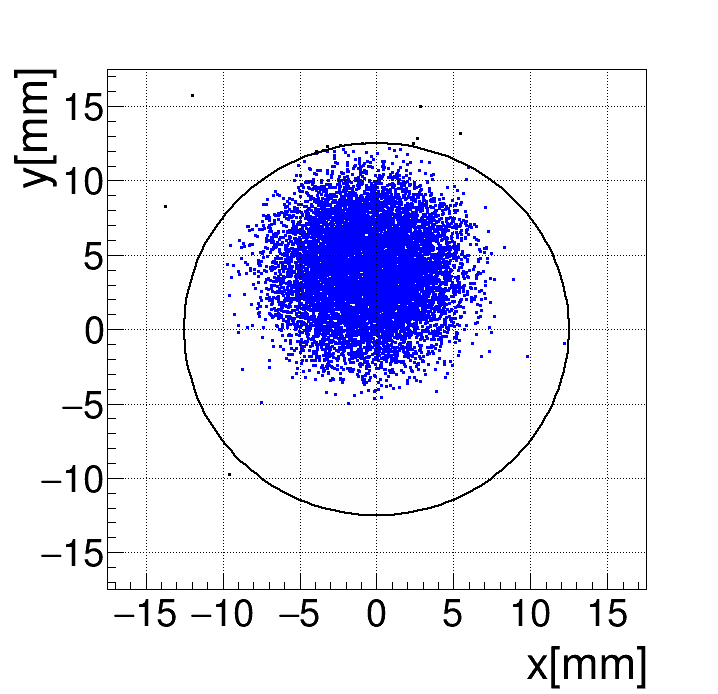}
}
\subfloat[\centering simulated beam profile at MCP3.5]{%
  \includegraphics[width=0.28\textwidth]{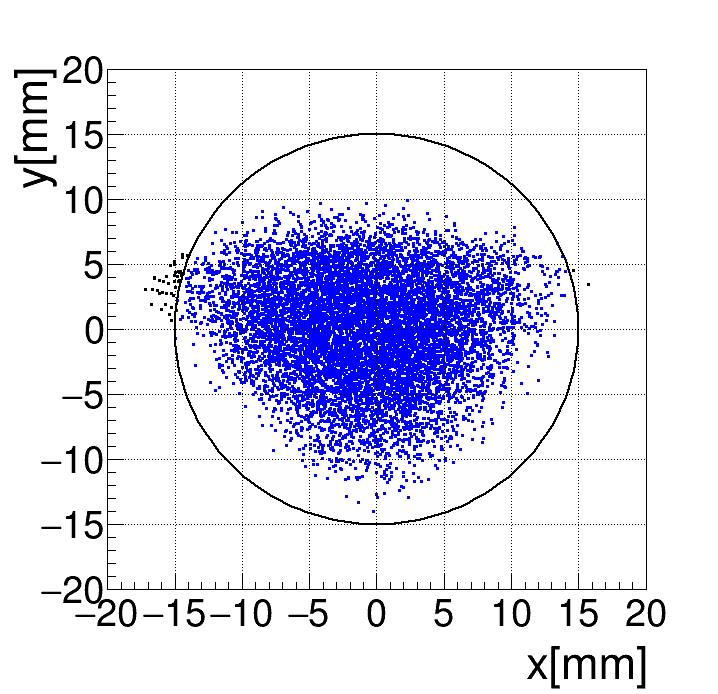}
}
\subfloat[\centering Linear correlation between initial plasma radius and beam size (FWHM) at MCP3.]{% empty placeholder
  \includegraphics[width=0.28\textwidth]{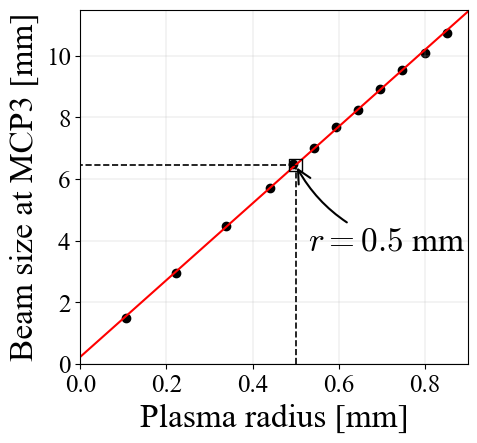}
}

\subfloat[\centering experimentally obtained beam profile at MCP3]{%
  \includegraphics[width=0.28\textwidth]{2024-11-08-0904_8MHz.png}
}
\subfloat[\centering experimentally obtained beam profile at MCP3.5]{%
  \includegraphics[width=0.28\textwidth]{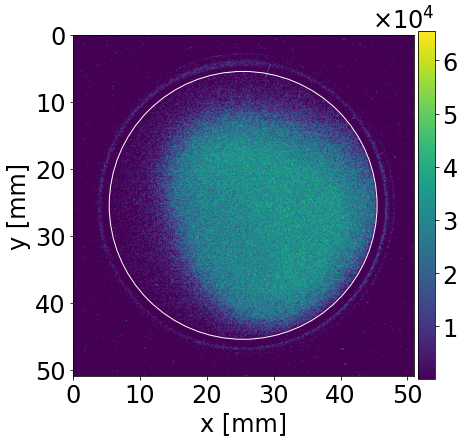}
}
\subfloat[\centering simulated beam profile at the target cavity]{
  \includegraphics[width=0.28\textwidth]{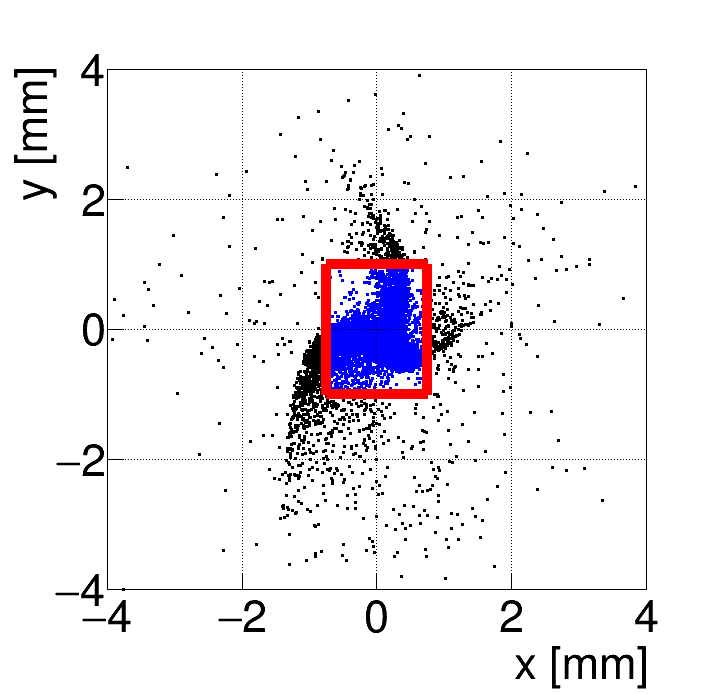}
}
\caption{ Simulated (a, b) and corresponding experimental (d, e) antiproton beam spatial distributions at MCP3 and MCP3.5.
(c) Linear correlation between the initial plasma radius and the beam size (FWHM) at MCP3, with simulation results (black dots) and a linear fit (red line). The initial plasma radius for the simulation was determined to be 0.5 mm by matching the experimental beam size with this linear fit. Simulated beam profile at the reaction target is drawn as (f). The red rectangle marks the cavity target window, and the blue dots represent particles passing through the cavity. }
\label{fig:simulation}
\end{figure}

A simulation study using WARP PIC (Particle-In-Cell),
a multi-dimensional particle simulation code especially developed for plasma and beam simulation~\cite{friedman},
was conducted to estimate the characteristics of the antiproton plasma and to optimize beam transport~\cite{Yoo_2022}.
The initial antiproton plasma is modeled as a uniformly distributed spheroid with zero kinetic energy.
Particle distributions with various plasma radii are generated, with the total number of particles fixed at $5 \times 10^6$, which corresponds to the antiproton count measured in the specific experimental configuration accumulated for comparison with the simulation.
The extraction sequence after the rotating wall compression is simulated,
using the experimental values for extraction and transport.
Starting from the spheroidal plasma confined in the trap, antiprotons are released downstream and accelerated by rapidly switching HV2 to high voltage.
The antiproton beam is then focused by four Einzel lenses, and the beam profile is recorded at MCP3 for various initial plasma radii.
As shown in Fig.~\ref{fig:simulation} (c), the initial plasma radius and the resulting beam radius at MCP3 exhibit a linear correlation. By comparing the simulation results (a) with experimental data (d), the initial plasma radius is estimated to be 0.5~mm.

Using the initial particle distribution simulated data with a radius of 0.5~mm, a further simulation was carried out to study the antiproton beam transport and to estimate the number of antiprotons passing through the target cavity.
As shown in Fig.~\ref{fig:simulation} (b) and (e), the simulated beam size and shape at MCP3.5, located midway between MCP3 and the target cavity, agree well with the experimental data. The simulation estimates the beam emittance to be 3~mm$\cdot$mrad.
The antiproton beam is subsequently transported through the target cavity. In the experiment, $9.98(0.34)\times10^5$ antiprotons are detected downstream of the target cavity, in good agreement with the simulated value of $1.06 \times 10^6$. Based on this validation, the beam optics were optimized via simulation to maximize the number of particles passing through the target cavity. The simulation predicts that $3.9\times10^6$ antiprotons pass through the target cavity, with the corresponding spatial distribution shown in Fig.~\ref{fig:simulation} (f).

These results demonstrate that the reaccelerated antiproton beam can be successfully transmitted through the narrow inlet ($1.5 \times 2.0$ mm$^2$) of the target cavity with a significantly higher flux compared to the previous beamline configuration without the antiproton trap~\cite{gbar_2022}. Consequently, the antihydrogen production rate is expected to increase, thereby facilitating the core experimental objectives of the GBAR experiment.

%%%%%%%%%%%%%%%%%%%%%%%%%%%%%%%%%%%%%%%%%%%%%%%%%%%%%%%%%%%%%
\section{Conclusion}
%%%%%%%%%%%%%%%%%%%%%%%%%%%%%%%%%%%%%%%%%%%%%%%%%%%%%%%%%%%%%

A unique antiproton beam preparation system comprising a pulsed drift-tube-based decelerator and a Penning-Malmberg trap, equipped with a cooled beam re-acceleration and bunching system, has been demonstrated using the antiproton beam provided by CERN's AD/ELENA facility.
We have reached a record trapping efficiency of 56(3)\% relative to the ELENA beam intensity. Furthermore, we optimized the antiproton trap capacity to accumulate $6.4(0.4) \times 10^7$ antiprotons by stacking 18 ELENA injections, representing the largest quantity of antiprotons reported to date. 
With this improved performance, the antihydrogen production rate will be significantly enhanced, providing an essential foundation for the future formation of a positive antihydrogen ion ($\mathrm{\overline{H}}^+$), which is the primary scientific objective of the GBAR experiment.

% A unique antiproton beam preparation system comprising a pulsed drift-tube-based decelerator and a Penning-Malmberg trap, equipped with a cooled beam re-acceleration and bunching system, has demonstrated its performance using the antiproton beam provided by CERN's AD/ELENA facility.
% The system has achieved a record efficiency of 56\% relative to the ELENA beam intensity. Furthermore, the antiproton trap demonstrated the capacity to accumulate $6.4(0.4) \times 10^7$ antiprotons by stacking 18 ELENA pulses, representing the largest population reported to date. 
% With this improved performance, the antihydrogen production rate is significantly enhanced, thereby facilitating the primary scientific objectives of the GBAR experiment.

%%%%%%%%%%%%%%%%%%%%%%%%%%%%%%%%%%%%%%%%%%%%%%%%%%%%%%%%%%%%%
\section*{Acknowledgements}
%%%%%%%%%%%%%%%%%%%%%%%%%%%%%%%%%%%%%%%%%%%%%%%%%%%%%%%%%%%%%

We thank L. Ponce and the AD/ELENA team as well as F. Butin and the CERN EN team for their fruitful collaboration.
We also thank A. Beynel for his continued support with the alignment of the trap.
%A. Prost and T. Stadlbauer from CERN and D. C. Faircloth from RAL are also warmly thanked for their help on HV techniques for the decelerator as well as A. Sinturel for help and expertise on vacuum.
%We recall the help of A. Leite and A. Husson for their work during the commissioning phase of the experiment. 
This work is supported by: JSPS KAKENHI Grant-in-Aid for Scientific Research A 20H00150 and Fostering Joint International Research A 20KK0305 (Japan), 
%ANTION ANR-14-CE33-0008 (France), 
Programme National Gravitation, Références, Astronomie, Métrologie (PNGRAM), the INP and IN2P3, CNRS, SPHINX ANR-22-CE31-0019 (France), the Swiss National Science Foundation (Switzerland) grants 197346, 216673 and 232699 and ETH Zurich (Switzerland) grant ETH-46 17-1, the Swedish Research Council (VR) grants 2017-03822 and 2021-04005 and the following grants from Korea: IBSR016-Y1, IBS-R016-D1, UBSI Research Fund (No. 1.220116.01) of UNIST, POSTECH Initial Settlement Support Fund, NRF-2016R1A5A1013277, NRF-RS-2022-00143178, NRF-2021R1A2C3010989, and NRF-2016R1A6A3A11932936. The GBAR collaboration is an International Research Network, supported by CNRS, France.

% We thank F. Butin and the CERN EN team, L. Ponce and the AD/ELENA team for their fruitful collaboration. A. Prost and T. Stadlbauer from CERN and D. C. Faircloth from RAL are also warmly thanked for their help on HV techniques for the decelerator as well as A. Sinturel for help and expertise on vacuum.
% We recall the help of A. Leite and A. Husson for their work during the commissioning phase of the experiment. This work is supported by: JSPS KAKENHI Grant-in-Aid for Scientific Research A 20H00150 and Fostering Joint International Research A 20KK0305 (Japan), ANTION ANR-14-CE33-0008 (France), Programme National Gravitation, Références, Astronomie, Métrologie (PNGRAM), Institut de Physique, CNRS (France), the Swiss National Science Foundation (Switzerland) grants 197346 and 216673 and ETH Zurich (Switzerland) grant ETH-46 17-1, the Swedish Research Council (VR) grants 2017-03822 and 2021-04005 and the following grants from Korea: IBSR016-Y1, IBS-R016-D1, UBSI Research Fund (No. 1.220116.01) of UNIST, NRF-2016R1A5A1013277, NRF-RS-2022-00143178, NRF-2021R1A2C3010989, and NRF-2016R1A6A3A11932936. The GBAR collaboration is an International Research Network, supported by CNRS, France.

% %%% Line Numbers for editors
% \nolinenumbers

%%%%%%%%%%%%%%%%%%%%%%%%%%%%%%%%%%%%%%%%%%%%%%%%%%%%%%%%%%%%%
% \section*{References}
%%%%%%%%%%%%%%%%%%%%%%%%%%%%%%%%%%%%%%%%%%%%%%%%%%%%%%%%%%%%%
\bibliographystyle{unsrt}
\bibliography{reference}

\end{document}